\documentclass[EPiC]{easychair}

\usepackage{amsmath}

\title{Some Experiments with Twee-Style Goal-Directedness}

\author{Stephan Schulz}

\institute{
  DHBW Stuttgart\\
  Stuttgart, Germany\\
  \email{schulz@eprover.org}
}

\authorrunning{Schulz}

\titlerunning{Goal-Directedness}

\newcommand{\mw}[1]{\ensuremath{\operatorname{#1}}}

\begin{document}

\maketitle

\begin{abstract}
  In saturation-based theorem proving, selecting the next clause for
  processing is a major concern. Twee has successfully applied the
  idea of preferring clauses that share terms with the conjecture by
  adding equational definitions to transform the problem. In this
  paper, we apply the idea to the full first-order case, and offer an
  alternative implementation based on shared terms. Both approaches
  have complementary applications and show very promising results.
\end{abstract}

\section{Introduction}

Automated theorem proving tries to automatically find formal proofs
for logical statements. The most well-developed subfield deals with
problems in first-order logic (with equality), but there now are also
provers for higher-order logics. An important sub-class are provers
for pure unit-equational logic. The predominant paradigm for current
high-performance provers is saturation. In this approach, the axioms
and the negated conjecture are converted into a flat set of clauses
that is unsatisfiable if and only if the original conjecture
holds. These clauses are then combined using sound inference rules
until the empty clause, as an explicit witness of unsatisfiability,
has been derived. For this standard saturation approach, the calculus
deals with a flat set of clauses - in particular, the calculus itself
does not distinguish between clauses created from the goal and clauses
created from the axioms.

However, for most situation we can assume that the axiomatization is
consistent, and hence that any derivation of the empty clause must
involve at least one clause from the conjecture. Intuitively, taking
this fact into account should help us to find proofs faster. In
particular, it seems promising to not only prefer pure goal clauses,
but also clauses that share properties with goal clauses, and hence
may constructively interact with them.

In this paper, we generalise a problem transformation technique
originally introduced by the equational prover
Twee~\cite{Smallbone:CADE-2021} to the full first-order (and
higher-order) case. We also propose and implement a different solution
based on the same idea, utilizing the shared term
paradigm~\cite{Schulz:IWIL-2025} at the heart of the prover
E~\cite{SCV:CADE-2019}, that works without a real problem
transformation and hence can be applied more selectively and in
combination with other search strategies.

\section{Goal-directed Proof Search}

As mentioned above, modern saturation calculi like e.g.\
Superposition~\cite{BG94} reduce the theorem proving problem to the
unsatisfiability of a flat clause set. There are also calculi that try
to accommodate knowledge about the conjecture status of a clause at
the calculus level, e.g.\ Model
Elimination~\cite{Loveland:JACM-68,LS:Handbook-2001} and
Set-of-Support Resolution~\cite{WRC:JACM-65}. However, these are not
compatible with the ordering-based redundancy criteria and especially
the rewrite-based treatment of equality that are are critical for the
success of state-of-the-art provers.

We can, however, use information about the goal-status of clauses for
heuristic control of the proof search. Most leading provers use a
variant of the \emph{given-clause algorithm}, i.e.\ they split the
proof state into two sets, \emph{processed} and \emph{unprocessed}
clauses, and heuristically pick clauses from the unprocessed set to
interact with previously processed clauses. The order of selection of
the so-called \emph{given clause} is generally considered the most
important choice point for saturation-based provers. It is typically
based on clause size (determined by some form of symbol counting), and
clause age. Smaller and older clauses are
preferred~\cite{SM:IJCAR-2016}, with different provers supporting
different ways to instantiate and interleave clause evaluation
functions. For completeness, clause selection should be fair, i.e.\ no
non-redundant clause should be delayed forever. Typically, this is
achieved by making sure that no clause is delayed forever, which can
e.g.\ be achieved by occasionally picking the oldest clause, or, if
the system supports subsumption, using at least one symbol-counting
heuristic that ensures that for each size limit there are at most a
finite number of $\alpha$-equivalent clauses under that size.

Our own prover, E, has long supported setting up an arbitrary number
of priority queues, each sorted by a different clause evaluation
function, and also a large set of generic clause evaluation functions
that can be instantiated per queue. Clauses are then drawn from each
queue in a weighted manner. For example, the specification
\texttt{(10.Clauseweight(ConstPrio,1,1,1), 1.FIFOWeight(ConstPrio))}
sets up two clause queues, with the first being ordered by symbol
count, the second being ordered by clause age, and 10 out of every 11
clauses picked from the first, the last one from the second queue. In
addition to plain symbol counting, evaluation functions in E can also
take the term ordering into account (typically giving a higher weight
to maximal literals and/or maximal terms in them), give different
weight to symbols based on type, or let the user select particular
weights for individual symbols. In addition, the system supports
hints~\cite{Veroff:JAR-1996,GJSU:ITP-2018}, i.e.\ the provision of a
\emph{watch list} of clauses that should be preferentially processed
if encountered.

E also supports giving preference to goal-related clauses, either
using hard priority functions preferring Set-of-Support-clauses, or
clause evaluation functions that give a lower (i.e. better) weight to
clauses containing goal symbols. These approaches contribute to mixed
clause selection strategies, and are valuable in portfolios, but
overall have limited effect~\cite{SM:IJCAR-2016}.

Recently, Smallbone has suggested and implemented a goal-directed
transformation for his unit-equational prover
Twee~\cite{Smallbone:CADE-2021}. In particular, he introduces new
equational definitions that rewrite all ground terms from the
conjecture to simple constants. As a result, all clauses that
originally shared subterms with the conjecture will become smaller (in
the number of symbols), and will hence be preferred. He reports useful
results for the unit-equational case. However, while this method
achieves the desired effect of preferring clauses with goal terms, it
has the disadvantage that it adds more symbols and clauses to the
search space.

Here, we build on the basic idea, apply it to more expressive
logics, and suggest an alternative implementation of the idea that
avoids some of the disadvantages.

\section{Implementation}

\subsection{Explicit Definitions}

Our first implementation attempts to match Twee's explicit
transformation approach. After loading a problem specification, we run
preprocessing and clausification (if necessary) as usual, giving us
the problem as a flat set of clauses. During clausification, we track
which clauses are clausified axioms, and which represent clausified
conjectures. We then extract all ground terms from conjecture clauses
and add recursive equational definitions to flatten all the extracted
terms into (usually new) constants. Consider a signature with function
symbol $f/2$ and constants $a/0, b/0$, and the conjecture clause
$f(f(a,b),b) \not\simeq f(b,f(b,a))$. The non-constant conjecture
ground terms are:
$$\{f(f(a,b),b), f(a,b), f(b,f(b,a)), f(b,a)\}$$

\noindent
We would then add the following new definitions (where the $c_i/0$ are
fresh constants):
\begin{itemize}
\item $f(a,b) \simeq c_1$
\item $f(c_1,b) \simeq c_2$
\item $f(b,a) \simeq c_3$
\item $f(b,c_3) \simeq c_4$
\end{itemize}

Note that these equations allow the rewriting of all ground conjecture
subterms to a normalform consisting of a single constant. To activate
this transformation, the user can set the global command line options
\texttt{--goal-defs --goal-subterm-defs}.

As an alternative, we also implemented a version in which only
definitions for the maximal (with respect to the subterm ordering)
ground conjecture subterms receive this treatment. In our example,
that would add the following definitions:

\begin{itemize}
\item $f(f(a,b),b) \simeq c_1$
\item $f(b,f(b,a)) \simeq c_2$
\end{itemize}

This version is selected by just passing \texttt{--goal-defs} to the
prover.

One important detail is that we do not aggressively and automatically
apply the generated definitions, but only add the new definitions to
the axiom set. Since these clauses are very small, they are typically
selected early in the proof search, and hence will be used to simplify
all newly generated and all selected given-clauses.

\subsection{...but Laziness is a Cardinal Virtue}

The main purpose of Twee's transformation is to make conjecture ground
terms syntactically small, so that symbol-counting heuristics prefer
clauses in which they occur. However, this comes at the cost of adding
additional clauses and additional symbols to the problem, increasing
the size of the search space. Also, clauses first need to be
explicitly rewritten for the transformation to take effect.

In E, we found an alternative approach to achieve the same result
without an explicit syntactic transformation. E is built around the
idea of shared terms. All persistent terms are stored in a \emph{term
  bank}, and each term is represented exactly once in this term
bank~\cite{Schulz:IWIL-2025}. For our alternative implementation of
goal-directedness, we use this feature. Terms that occur in conjecture
clauses are simply marked in the term bank. We implement an
alternative symbol-counting clause evaluation function that checks for
these flags and evaluates conjecture ground terms differently.

The default symbol counting function can be recursively defined by
\[
\begin{array}{ll}
  \mw{sc}(X, w_{v}, w_{fun})= w_{v} &\\[1.5ex]
  \mw{sc}(f(t_1,\ldots, t_n), w_{v}, w_{fun}) =
  w_{fun}+\sum_{1\leq i\leq n} \mw{sc}(t_i, w_{v}, w_{fun}) & \\
\end{array}
\]

where $X$ is an arbitrary variable, and $f/n$ is an arbitrary function
symbol. Note that the second case includes the case of $n=0$, i.e.\
$f$ being a constant. The two (non-negative integer) parameters
$w_{v}$ and $w_{fun}$ allow us to give different weights to function
symbols and variable occurrences.

Taking goal ground terms into account, the new goal-directed term
weight is defined as follows:

\small
\[
  \begin{array}{ll}
    \mw{gd}(X, w_{v}, w_{fun}, c_{gt}, m_{gt}) & = w_{v} \\[1.5ex]
    \mw{gd}(f(t_1,\ldots, t_n), w_{v}, w_{fun}, c_{gt}, m_{gt}) & =
    \begin{cases}
      c_{gt} + m_{gt}\cdot \mw{sc}(f(t_1,\ldots t_n), w_{v},
      w_{fun}) & \text{if }f(t_1,\ldots t_n) \\
               & \text{is a goal } \\
               & \text{ground term}\\
      w_{fun}+\sum_{1\leq i\leq n} \mw{gd}(t_i, w_{v}, w_{fun}, c_{gt},
      m_{gt})  & \text{otherwise}\\
    \end{cases}\\
  \end{array}
\]
\normalsize

For terms that contain no conjecture ground terms, the result is the
same as for plain symbol counting. However, if a conjecture ground
term is encountered, the standard symbol counting weight is modified
by a constant (integer) offset and a (real) multiplier (the two new
parameters $c_{gt}$ and $m_{gt}$). For $c_{gt} = 0$ and $m_{gt}=1$,
the function is equivalent to plain symbol counting. For $m_{gt}=0$,
$c_{gt} = w_{fun}$, the complete conjecture ground term is counted as
a single constant, as it would be after the explicit goal-directed
transformation described in the previous section. However, the
parameters allow a whole spectrum of possible alternative ways to
weight the conjecture ground term, including, with $m_{gt}=0$ and
$c_{gt} = 0$, not at all. Since there are only finitely many
conjecture ground terms, this does not compromise fairness as long as
$m_{gt}\geq 0$ and $c_{gt}\geq 0$.

We have implemented this new term evaluation function as and lifted it
to a heuristic clause evaluation functions: \texttt{GDWeight(prio\_fun,
  $w_f$, $w_v$, $m_{pos}$, $m_{gd}$, $c_{gt}$)}, where
\texttt{prio\_fun} is a priority function (allowing a high-level
classification of clauses by various classes), $m_{pos}$ is a
multiplier applied to the weight of positive literals, and the other
parameters are as described above.

\section{Experimental Results}

We performed several experiments to compare the performance of the
goal-directed transformation and evaluation function. In all cases, we
picked all typed and untyped first-order problems from
TPTP~\cite{Sutcliffe:JAR-2017}, version~9.2.1, and ran them on the
Miami instance of the StarExec cluster~\cite{SST:IJCAR-2014}. The
computers are equipped with 128GB of memory and Intel Xeon E5-2620
processors running at 2.10GHz, with one job scheduled per CPU. The
operating system is Ubuntu 24.04.3~LTS, running the Linux Kernel
6.8.0-71-generic. All problems were run with a wall clock limit of 90s
and a CPU limit of 60s.

There is a total of 19351 problems in our set, of which 1455 are
unit-equational problems (UEQ), 6889 are non-unit CNF problems, 9091
are untyped full first-order problems (FOF), and 1916 are typed
first-order problems (TF0).

The version of E used in the experiments is available from GitHub
under~\url{https://github.com/eprover/eprover} (Tag E-3.5.1), the new
extensions will be included in the upcoming release of E~3.5.

Based on~\cite{SM:IJCAR-2016}, we used
\texttt{(10.Clauseweight(ConstPrio,1,1,1),1.FIFO\-Weight(Const\-Prio))}
as the baseline clause selection heuristic, i.e.\ a strategy that
repeatedly picks 10 clauses according to symbol counting (with equal
weight for variables and function symbols), then one clause based on
age. This was the strongest symbol-counting/fifo heuristic in this
previous work.

The other search parameters were set as in our previous work and
represent one of the strongest general-purpose settings we have found
so far.\footnote{The exact settings are \texttt{--proof-object
    --simul-paramod --forward-context-sr --strong\--des\-tructive-er
    --destructive-er-aggressive --destructive-er -tKBO6 -winvfreqrank
    -c1 -Ginvfreq -F1 -WSelectMaxLComplexAvoidPosPred
    -H(10.Clauseweight(ConstPrio,1,1,1),1.FIFOWeight(Const\-Prio))
    --Cpu-limit=60}.}

\begin{table}[th]
  \centering
  \small
  \begin{tabular}{|l|rrr|rrr|rrr|rrr|}
  \hline
  & \multicolumn{3}{|c|}{Symbol Counting}
  & \multicolumn{3}{|c|}{Recursive GDT}
  & \multicolumn{3}{|c|}{Top level GDT}
  & \multicolumn{3}{|c|}{GD-Evaluation}\\
  \hline
  & Prfs & Sats & Succ & Prfs & Sats & Succ & Prfs & Sats & Succ & Prfs & Sats & Succ\\
  \hline
  UEQ &  744 &   97 &  841  &  819 &   97 &  916  &  776 &   97 &  873  &  761 &   97 &  858 \\
  CNF & 3922 &  529 & 4451  & 4331 &  537 & 4868  & 3964 &  528 & 4492  & 4134 &  529 & 4663 \\
  FOF & 3484 &  413 & 3897  & 3845 &  414 & 4259  & 3558 &  413 & 3971  & 3674 &  412 & 4086 \\
  TF0 &  221 &   31 &  252  &  228 &   31 &  259  &  222 &   31 &  253  &  230 &   31 &  261 \\
  \hline
  ALL & 7627 &  973 & 8600  & 8404 &  982 & 9386  & 7744 &  972 & 8716   & 8038 &  972 & 9010 \\
  \hline
  \end{tabular}
  \normalsize
\caption{Number of successes for different strategies and problem
    classes}
  \label{tab:results}
\end{table}

\begin{figure}
  \centering
  \includegraphics[width=0.8\textwidth]{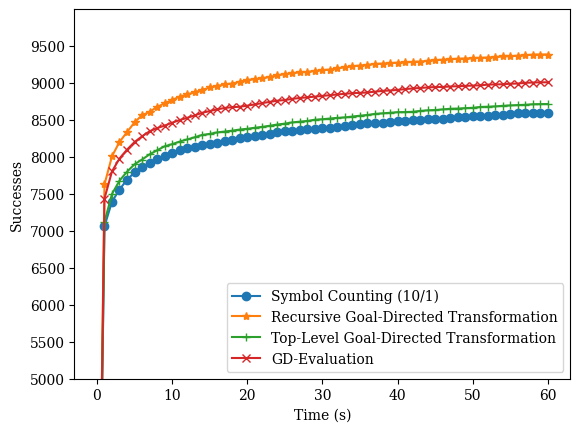}
  \caption{Successes over time for different strategies }
  \label{fig:basic_res}
\end{figure}

Table~\ref{tab:results} summarizes the results of our experiments. The
four major columns represent the different search strategies, with
\emph{Symbol counting} being the standard strategy described
above. \emph{Recursive GDT} shows the results with the full Twee-style
goal-directed transformation enabled, adding new rewrite definitions
for all subterms (options \texttt{--goal-defs
  --goal-subterm-defs}). The column \emph{Top level GDT} lists results
with definitions just added for the subterm-maximal ground terms in
conjecture clauses. Finally, \emph{GD-Evaluation} replaces
\texttt{Clauseweight} with the new goal-directed evaluation function,
parameterized to model the recursive goal-directed transformation,
i.e.\ setting $m_{gt}=0$ and $c_{gt}=1$:
\texttt{(10.GDWeight(ConstPrio,1,1,1,0,1),1.FIFO\-Weight(ConstPrio))}. For
each strategy, we count proofs found (\emph{Prfs}),
(counter-)saturations found (\emph{Sats}) and total successes
(\emph{Succ}). Figure~\ref{fig:basic_res} visualises the number of
successes over time.

First, looking at all problem types and all successes, we can see that
the baseline strategy solves 8600 problems. Applying the recursive
goal-directed transformation increases this to 9386 solutions, a quite
significant increase of nearly 800 solutions. Just adding definitions
for the top-terms, on the other hand, has only the relatively minor
effect of increasing solutions to 8716. Our shared-term-based
evaluation function, finally, solves 9010 problems, about 400 problems
over baseline. Both the successes of the explicit and implicit
goal-directed transformation/heuristic are quite significant, and
something rarely found with the first set of experiments. However, the
stronger result of the explicit transformation indicates that there
probably is another effect beyond the heuristic selection at
work. Looking at some problems, we do see that sometimes superposition
inferences with added definitions create useful instantiations for the
proof search.

Looking at the more detailed results, we can see that the relative
performance increase is very similar across all 4 classes
(unit-equational, where the technique originated, clause normal form,
full first-order and typed first-order). On the other hand, nearly all
of the gains are made for provable problems, with only a very minor
effect on counter-saturations.

We also observe that these strategies are still quite
complementary. All 4 strategies together solve 9804 problems (8820
proofs, 984 saturations), about 1200 problems more than the baseline,
and about 500 problems more than the best strategy.

With this initial set of experiments, we do not aggressively apply the
goal-directed transformation immediately, but only add the rewrite
rules to the clause set.

\begin{figure}
  \centering
  \includegraphics[width=0.5\textwidth]{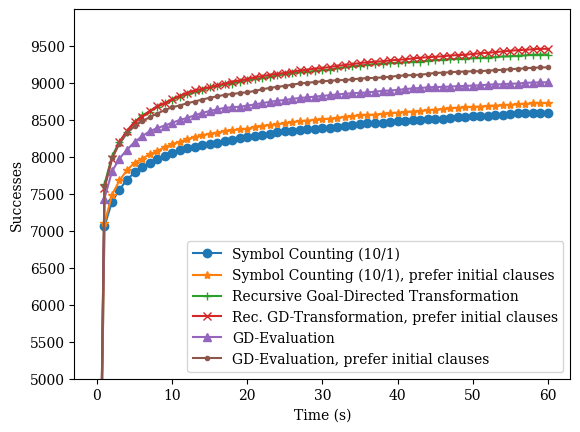}%
  \includegraphics[width=0.5\textwidth]{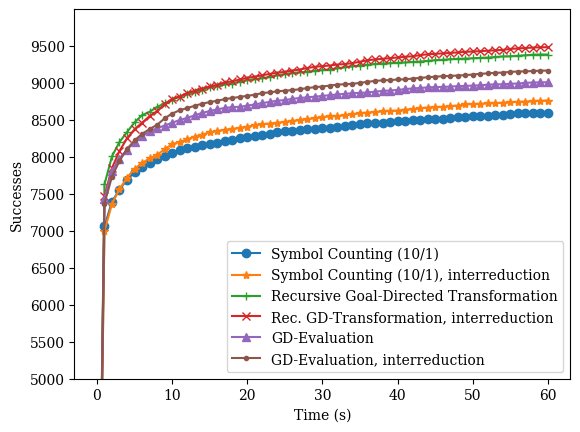}
  \caption{Successes over time with preferred processing of initial
    clauses (left) and early interreduction (right)}
  \label{fig:res2}
\end{figure}

There are two options of E that force all or most of the
simplifications to be done before going into deep proof-search. First,
E can be configured to process all initial clauses before any of the
newly generated ones (option \texttt{--prefer-initial-clauses}). And
secondly, E can be configured to perform a complete interreduction of
the input clauses before starting the full given-clause search
procedure (option \texttt{--presaturation-interreduction}). Both of
these options are often, but not universally helpful. We have
performed the experiments for the full recursive transformation and
for the goal-directed clause evaluation
function. Table~\ref{tab:results2} summarizes the results if we enable
these options. The data is visualised (with the addition of symbol
counting for better comparison) in Figure~\ref{fig:res2}.

\begin{table}[th]
  \centering
  \small
  \begin{tabular}{|l|rrr|rrr|rrr|rrr|}
  \hline
  & \multicolumn{3}{|c|}{Rec.\ GDT (pic)}
  & \multicolumn{3}{|c|}{GD-Evaluation (pic)}
  & \multicolumn{3}{|c|}{Rec.\ GDT (ir)}
  & \multicolumn{3}{|c|}{GD-Evaluation (ir)}\\
  \hline
  & Prfs & Sats & Succ & Prfs & Sats & Succ & Prfs & Sats & Succ & Prfs & Sats & Succ\\
  \hline
   UEQ &  823 &   97 &  920  &  756 &   97 &  853  &  818 &   97 &  915  &  761 &   97 &  858 \\
   CNF & 4372 &  537 & 4909  & 4243 &  530 & 4773  & 4314 &  538 & 4852  & 4169 &  528 & 4697 \\
   FOF & 3884 &  410 & 4294  & 3760 &  414 & 4174  & 3964 &  412 & 4376  & 3802 &  413 & 4215 \\
   TF0 &  229 &   31 &  260  &  234 &   31 &  265  &  226 &   31 &  257  &  226 &   31 &  257 \\
  \hline
   ALL & 8485 &  978 & 9463  & 8237 &  975 & 9212  & 8504 &  981 & 9485  & 8197 &  972 & 9169 \\
  \hline
  \end{tabular}
  \normalsize
\caption{Number of successes for strategies preferring initial clauses
  (pic) or using pre-interreduction (ir)}
  \label{tab:results2}
\end{table}

While both options increase the number of successes for all
heuristics, we observe that the effect of both options is fairly small
for the explicit goal-directed transformation (where we expected a
bigger effect) and significantly larger for the purely heuristic
approach. In particular, the explicit transformation gains about 80
solutions when preferring initial clauses and about 100 successes with
early interreduction, while the heuristic approach gains about 200 and
160, respectively.

Combining all 8 different strategies discussed here, E solves 10216
problems, including 15 problems with TPTP rating 1.0, i.e.\ previously
unsolved problems.

\section{Future Work}

The first step is, of course, to explore more of the parameter space,
and to see if there are additional low-hanging fruits, both by varying
the parameter of the search, and by combining the goal-directed
transformations and evaluation functions with a wider range of
existing heuristics.

The explicit transformations can be combined with any and all existing
strategies. However, because they modify the actual clause set, it is
not possible to use them in mixed settings - it constitutes an
all-or-nothing proposition. The goal-directed evaluation function can,
of course, be combined into mixed selection strategies, but it does
not support all the various features of E's wide range of evaluation
functions. It may be necessary to implement goal-directed versions of
the most significant other evaluation functions. Alternatively, we can
try to synthesize the experience gained over time and develop a single
evaluation function with a broader range of parameters that can
simulate other existing functions with and without the goal-directed
aspect.

Finally, the Twee team has recently generalized the approach to add
definitions for non-ground terms (called \emph{abstractions}), but to
limit such abstractions to certain frequently useful
patterns~\cite{AJS:IJCAR-2026}. This approach does not quite as easily
lend itself to the shared-term paradigm, but could also be explored in
the context of richer logics.

\section{Conclusion}

The simple idea of preferring clauses that share ground terms with the
original conjecture has turned out to be surprisingly effective. The
future will show how far this idea will carry, and how well this
approach will interact in more complex heuristics. But even now we can
say that it is one of the techniques with a very good implementation
effort to performance gain ratio.

\bibliographystyle{alpha}
\bibliography{stsbib}

\end{document}